\documentclass{article}%
\usepackage{amsmath}
\usepackage{amsfonts}
\usepackage{amssymb}
\usepackage{graphicx}
\usepackage{fullpage}%
\begin{document}

\title{Are Quantum States Exponentially Long Vectors?}
\author{Scott Aaronson\thanks{Currently at the University of Waterloo and supported by
ARDA. \ Email: aaronson@ias.edu. \ This abstract is mostly based on work done
while I was a student at UC\ Berkeley, supported by an NSF\ Graduate
Fellowship.}}
\date{}
\maketitle

I'm grateful to Oded Goldreich for inviting me to the 2005 Oberwolfach Meeting
on Complexity Theory. \ In this extended abstract, which is based on a talk
that I gave there, I demonstrate that gratitude by explaining why Goldreich's
views about quantum computing are wrong.

Why should anyone care? \ Because in my opinion, Goldreich, along with Leonid
Levin \cite{levin:qc} and other \textquotedblleft extreme\textquotedblright%
\ quantum computing skeptics, deserves credit for focusing attention on the
key issues, the ones that ought to motivate quantum computing research in the
first place. \ Personally, I have never lain awake at night yearning for the
factors of a 1024-bit RSA integer, let alone the class group of a number
field. \ The real reason to study quantum computing is not to learn other
people's secrets, but to unravel the ultimate Secret of Secrets: \textit{is
our universe a polynomial or an exponential place?}

Last year Goldreich \cite{goldreich:qc}\ came down firmly on the
\textquotedblleft polynomial\textquotedblright\ side, in a short essay
expressing his belief that quantum computing is impossible not only in
practice but also in principle:

\begin{quotation}
As far as I am concern[ed], the QC model consists of exponentially-long
vectors (possible configurations) and some \textquotedblleft
uniform\textquotedblright\ (or \textquotedblleft simple\textquotedblright)
operations (computation steps) on such vectors \ldots\ The key point is that
the associated complexity measure postulates that each such operation can be
effected at unit cost (or unit time). \ My main concern is with this
postulate. \ My own intuition is that the cost of such an operation or of
maintaining such vectors should be linearly related to the amount of
\textquotedblleft non-degeneracy\textquotedblright\ of these vectors, where
the \textquotedblleft non-degeneracy\textquotedblright\ may vary from a
constant to linear in the length of the vector (depending on the vector).
\ Needless to say, I am not suggesting a concrete definition of
\textquotedblleft non-degeneracy,\textquotedblright\ I am merely conjecturing
that such exists and that it capture[s] the inherent cost of the computation.
\end{quotation}

My response consists of two theorem-encrusted prongs:\footnote{Sanjeev Arora
asked why I don't have \textit{three} prongs, thereby forming a $\psi$-shaped
pitchfork.} first, that you'd have trouble explaining even current
experiments, if you didn't think that quantum states really\textit{ were}
exponentially long vectors; and second, that for most complexity-theoretic
purposes, the exponentiality of quantum states is not that much
\textquotedblleft worse\textquotedblright\ than the exponentiality of
classical probability distributions, which nobody complains about. \ The first
prong is based on my paper \textquotedblleft Multilinear Formulas and
Skepticism of Quantum Computing\textquotedblright\ \cite{aar:mlin}; the second
is based on my paper \textquotedblleft Limitations of Quantum Advice and
One-Way Communication\textquotedblright\ \cite{aar:adv}.

\section*{Prong 1: Quantum States \textit{Are} Exponential\label{PRONG1}}

For me, the main weakness in the arguments of quantum computing skeptics has
always been their failure to suggest an answer to the following question:
\textit{what criterion separates the quantum states we're sure we can prepare,
from the states that arise in Shor's factoring algorithm?} \ I call such a
criterion a \textquotedblleft Sure/Shor separator.\textquotedblright\ \ To be
clear, I'm not asking for a red line partitioning Hilbert space into two
regions, \textquotedblleft accessible\textquotedblright\ and \textquotedblleft
inaccessible.\textquotedblright\ \ But a skeptic could at least propose a
complexity measure for quantum states, and then declare that a state of $n$
qubits is \textquotedblleft efficiently accessible\textquotedblright\ only if
its complexity is upper-bounded by a small polynomial in $n$.

In his essay \cite{goldreich:qc}, Goldreich agrees that such a Sure/Shor
separator would be desirable, but avers that it's not his job to propose one:

\begin{quotation}
My main disagreement with Scott is conceptual: He says that it is up to the
\textquotedblleft skeptics\textquotedblright\ to make a [concrete] suggestion
(of such a \textquotedblleft complexity\textquotedblright) and views their
[arguments] as weak without such a suggestion. \ In contrast, I think it is
enough for the \textquotedblleft skeptics\textquotedblright\ to point out that
there is no basis to the (over-simplified and counter-intuitive to my taste)
speculation by which a QC can manipulate or maintain such huge objects
\textquotedblleft free of cost\textquotedblright\ (i.e., at unit cost).
\end{quotation}

Motivated by the \textquotedblleft hands-off\textquotedblright\ approach of
Goldreich and other skeptics, in\ \cite{aar:mlin} I tried to carry out the
skeptics' research program for them, by proposing and analyzing possible
Sure/Shor separators. \ The goal was to illustrate what a scientific argument
against quantum computing might look like.

For starters, such an argument would be careful to assert the impossibility
only of \textit{future} experiments, not experiments that have already been
done. \ As an example, it would not dismiss exponentially-small amplitudes as
physically meaningless,\ since one can easily produce such amplitudes by
polarizing $n$\ photons each at $45^{\circ}$. \ Nor would it appeal to the
\textquotedblleft absurd\textquotedblright\ number of particles that a quantum
computer would need to maintain in coherent superposition---since among other
examples, the Zeilinger group's $C_{60}$\ double-slit experiments
\cite{arndt}\ have already demonstrated\ \textquotedblleft Schr\"{o}dinger
cat\ states,\textquotedblright\ of the form $\frac{\left\vert 0\right\rangle
^{\otimes n}+\left\vert 1\right\rangle ^{\otimes n}}{\sqrt{2}}$, for $n$ large
enough to be interesting for quantum computation.

Of course, the real problem is that, once we accept $\left\vert \psi
\right\rangle $\ and $\left\vert \varphi\right\rangle $\ into our set of
possible states, consistency almost \textit{forces} us to accept
$\alpha\left\vert \psi\right\rangle +\beta\left\vert \varphi\right\rangle $
and $\left\vert \psi\right\rangle \otimes\left\vert \varphi\right\rangle $\ as
well. \ So is there any defensible place to draw a line? \ This conundrum is
what led me to investigate \textquotedblleft tree states\textquotedblright:
the class of $n$-qubit pure states that are expressible by polynomial-size
\textit{trees} of linear combinations and tensor products. \ As an example,
the state%
\[
\left(  \frac{\left\vert 0\right\rangle +\left\vert 1\right\rangle }{\sqrt{2}%
}\right)  \otimes\cdots\otimes\left(  \frac{\left\vert 0\right\rangle
+\left\vert 1\right\rangle }{\sqrt{2}}\right)
\]
is a tree state. \ For that matter, so is \textit{any} state that can be
written succinctly in the Dirac ket notation, using only the symbols
$\left\vert 0\right\rangle ,\left\vert 1\right\rangle ,+,\otimes
,(,)$\ together with constants (no $\sum$'s are allowed). \ In evaluating tree
states as a possible Sure/Shor separator, we need to address two questions:
first, should all quantum states that arise in present-day experiments be seen
as tree states? \ And second, would a quantum computer permit the creation of
non-tree states?

My results imply a positive answer to the second question: not only could a
quantum computer efficiently generate non-tree states, but such states arise
naturally in several quantum algorithms.\footnote{On the other hand, I do not
know whether a quantum computer restricted to tree states always has an
efficient classical simulation. \ All I can show is that such a computer would
be simulable in $\mathsf{\Sigma}_{3}^{p}\cap\mathsf{\Pi}_{3}^{p}$, the third
level of the polynomial-time hierarchy.} \ In particular, let $C$ be a random
linear code over $\mathbb{GF}_{2}$. \ Then with overwhelming probability, a
uniform superposition over the codewords of $C$ cannot be represented by any
tree of size $n^{\varepsilon\log n}$, for some fixed $\varepsilon
>0$.\footnote{In Raz's proof, $\varepsilon$\ was about $10^{-6}$, but what's a
constant between friends (or more precisely, between theoretical computer
scientists)?} \ Indeed, $n^{\varepsilon\log n}$\ symbols would be needed even
to \textit{approximate} such a state well in $L_{2}$-distance, and even if we
replaced the random linear code by a certain explicit code (obtained by
concatenating the Reed-Solomon and Hadamard codes). \ I also showed an
$n^{\varepsilon\log n}$\ lower bound for the states arising in Shor's
algorithm, assuming an \textquotedblleft obviously true\textquotedblright\ but
apparently deep number-theoretic conjecture: basically, that the multiples of
a large prime number, when written in binary, constitute a decent erasure
code. \ All of these results rely on a spectacular recent advance in classical
computer science: the first superpolynomial lower bounds on \textquotedblleft
multilinear formula size,\textquotedblright\ which were proven by Ran Raz
\cite{raz}\ about a month before I needed them for my quantum application.
\ Incidentally, in all of the cases discussed above, I conjecture that the
actual tree sizes are exponential in $n$; currently, though, Raz's method can
only prove lower bounds of the form $n^{\varepsilon\log n}$.\footnote{I did
manage to prove an exponential lower bound, provided we restrict ourselves to
linear combinations $\alpha\left\vert \psi\right\rangle +\beta\left\vert
\varphi\right\rangle $\ that are \textquotedblleft manifestly
orthogonal\textquotedblright---which means that for all computational basis
states $\left\vert x\right\rangle $. either $\left\langle \psi|x\right\rangle
=0$ or $\left\langle \varphi|x\right\rangle =0$.}

Perhaps more relevant to physics, I also conjecture that 2-D and 3-D
\textquotedblleft cluster states\textquotedblright\ (informally, 2-D and 3-D
lattices of qubits with pairwise nearest-neighbor interactions)\ have
exponential tree sizes.\footnote{By contrast, I\ can show that 1-D cluster
states have tree size $O\left(  n^{4}\right)  $.} \ If true, this conjecture
suggests that states with enormous tree sizes might have already been observed
in condensed-matter experiments---for example, those of Ghosh et al.
\cite{grac}\ on long-range entanglement in magnetic salts. \ In my personal
fantasy land, once the evidence characterizing the ground states of these
condensed-matter systems became overwhelming, the skeptics would come back
with a \textit{new} Sure/Shor separator. \ Then the experimentalists would try
to refute \textit{that} separator, and so on. \ As a result, what started out
as a philosophical debate would gradually evolve into a scientific one---on
which progress not only can be made, but is.

\section*{Prong 2: It's Not That Bad\label{PRONG2}}

\textit{To describe a state of }$n$\textit{ particles, we need to write down
an exponentially long vector of exponentially small numbers, which themselves
vary continuously. \ Moreover, the instant we measure a particle, we
\textquotedblleft collapse\textquotedblright\ the vector that describes its
state---and not only that, but possibly the state of another particle on the
opposite side of the universe.} \ Quick, what theory have I just described?

The answer is classical probability theory. \ The moral is that, before we
throw up our hands over\ the \textquotedblleft extravagance\textquotedblright%
\ of the quantum worldview, we ought to ask: is it so much \textit{more}
extravagant than the classical probabilistic worldview? \ After all, both
involve linear transformations of exponentially long vectors that are not
directly observable. \ Both allow fault tolerance, in stark contrast with
analog computing. \ Neither lets us reliably pack $n+1$\ bits into an $n$-bit
state. \ And\ neither (apparently!) would provide enough power to solve
$\mathsf{NP}$-complete problems\ in polynomial time.

But none of this addresses the central complexity-theoretic question: if
someone gives you a polynomial-size quantum state, \textit{how much more}
\textit{useful} is that than being given a sample from a classical probability
distribution? \ In their textbook \cite{nc}, Nielsen and Chuang were getting
at this question when they made the following intriguing speculation:

\begin{quotation}
[W]e know that many systems in Nature \textquotedblleft
prefer\textquotedblright\ to sit in highly entangled states of many systems;
might it be possible to exploit this preference to obtain extra computational
power? \ It might be that having access to certain states allows particular
computations to be done much more easily than if we are constrained to start
in the computational basis.
\end{quotation}

To a complexity theorist like me, them's fightin' words---or at least,
complexity-class-definin' words. \ In particular, let's consider the class
$\mathsf{BQP/qpoly}$, which consists of all problems solvable in polynomial
time on a quantum computer, if the quantum computer has access to a
polynomial-size \textquotedblleft quantum advice state\textquotedblright%
\ $\left\vert \psi_{n}\right\rangle $\ that depends only on the input length
$n$. \ (For the uninitiated, $\mathsf{BQP}$\ stands for \textquotedblleft
Bounded-Error Quantum Polynomial-Time,\textquotedblright\ and $\mathsf{/qpoly}%
$\ means \textquotedblleft with polynomial-size quantum
advice.\textquotedblright) \ Note that $\left\vert \psi_{n}\right\rangle
$\ might be arbitrarily hard to prepare; for example, it might have the form
$2^{-n/2}\sum_{x}\left\vert x\right\rangle \left\vert f\left(  x\right)
\right\rangle $\ for an arbitrarily hard function $f$. \ We can imagine that
$\left\vert \psi_{n}\right\rangle $\ is given to us by a benevolent
wizard;\ the only downside is that the wizard doesn't know which input
$x\in\left\{  0,1\right\}  ^{n}$ we're going to get, and therefore needs to
give us a single advice state that works for all $x$.

The obvious question is this: \textit{is quantum advice more powerful than
classical advice?} \ In other words, does $\mathsf{BQP/qpoly}%
=\mathsf{BQP/poly}$, where $\mathsf{BQP/poly}$\ is the class of problems
solvable in quantum polynomial time with the aid of polynomial-size
\textit{classical} advice? \ As usual in complexity theory, the answer is that
we don't know. \ This raises a disturbing possibility: could quantum advice be
similar in power to \textit{exponential-size} classical advice, which would
let us solve any problem whatsoever (since we'd simply have to store every
possible answer in a giant lookup table)? \ In particular, could
$\mathsf{BQP/qpoly}$\ contain the $\mathsf{NP}$-complete problems, or the
halting problem, or even \textit{all} problems?

If you know me, you know I'd sooner accept that pigs can fly. \ But how to
support that conviction? \ In \cite{aar:adv}, I did so\ with the help of yet
another complexity class: $\mathsf{PostBQP}$, or $\mathsf{BQP}$\ with
postselection. \ This is the class of problems solvable in quantum polynomial
time, if at any stage you could measure a qubit and then \textit{postselect}
on the measurement outcome being $\left\vert 1\right\rangle $\ (in other
words, if you could kill yourself if the outcome was $\left\vert
0\right\rangle $, and then condition on remaining alive). \ My main result was
that $\mathsf{BQP/qpoly}$ is contained in $\mathsf{PostBQP/poly}$. \ Loosely
speaking, anything you can do with polynomial-size quantum advice, you can
also do with polynomial-size \textit{classical} advice, provided you're
willing to use exponentially more computation time (or settle for an
exponentially small probability of success).\footnote{In \cite{aar:pp} I
characterized $\mathsf{PostBQP}$\ exactly in terms of the classical complexity
class $\mathsf{PP}$ (Probabilistic Polynomial-Time),\ which consists of all
decision problems solvable in polynomial time by a randomized Turing machine,
which accepts with probability greater than $1/2$\ if and only if the answer
is \textquotedblleft yes.\textquotedblright\ \ Thus, a more conventional way
to state my result is $\mathsf{BQP/qpoly}\subseteq\mathsf{PP/poly}$.}

On the other hand, since $\mathsf{NP}\subseteq\mathsf{PP}$, this result still
says nothing about whether a quantum computer with quantum advice could solve
$\mathsf{NP}$-complete problems in polynomial time. \ To address that
question, in \cite{aar:adv}\ I also created a \textquotedblleft relativized
world\textquotedblright\ where $\mathsf{NP}\not \subset \mathsf{BQP/qpoly}$.
\ This means, roughly, that there is no \textquotedblleft
brute-force\textquotedblright\ method to solve $\mathsf{NP}$-complete problems
in quantum polynomial time, even with the help of quantum advice: any proof
that $\mathsf{NP}\subset\mathsf{BQP/qpoly}$\ would have to use techniques
radically unlike any we know today.

In my view, these results support the intuition that quantum states are
\textquotedblleft more like\textquotedblright\ probability distributions over
$n$-bit strings than like exponentially-long strings to which one has random
access. \ If exponentially-long strings were rocket fuel, and probability
distributions were grape juice, then quantum states would be wine---the
alcoholic \textquotedblleft kick\textquotedblright\ in this analogy being the
minus signs. \ I can imagine someone objecting: \textquotedblleft What a load
of nonsense! \ Whether quantum states are more like grape juice or rocket fuel
is not a mathematical question, about which theorems could say
anything!\textquotedblright\ \ To which I'd\ respond: if results such as
$\mathsf{BQP/qpoly}\subseteq\mathsf{PostBQP/poly}$, which sharply limit the
power of quantum advice, do not count as evidence against Goldreich's view of
quantum states, then what \textit{would} count as evidence? \ And if nothing
would count, then how scientifically meaningful is that view in the first place?

A meatier objection centers around a recent result of Raz \cite{raz:all}, that
a quantum interactive proof system where the verifier gets quantum advice can
solve any problem whatsoever---or in complexity language, that
$\mathsf{QIP/qpoly}$ equals $\mathsf{ALL}$. \ (Here $\mathsf{ALL}$\ is the
class of all problems, which means, literally, the class of all problems.) \ I
made a related observation in \cite{aar:adv}, where I\ pointed out that
$\mathsf{PostBQP/qpoly}$ equals $\mathsf{ALL}$.\footnote{Here's a one-sentence
proof: given the advice state $2^{-n/2}\sum_{x}\left\vert x\right\rangle
\left\vert f\left(  x\right)  \right\rangle $, to evaluate the Boolean
function $f$\ on any given input $x$ we simply need to measure in the standard
basis, then postselect on seeing the $\left\vert x\right\rangle $\ of
interest.} \ However, a key point about both results is that \textit{they have
nothing to do with quantum computing}, and indeed, would work just as well
with classical randomized advice. \ In other words, the classes
$\mathsf{IP/rpoly}$\ and $\mathsf{PP/rpoly}$\ are also equal to $\mathsf{ALL}%
$. \ Since I\ like making conjectures, I'll conjecture more generally that
quantum advice does not wreak much havoc in the complexity zoo that isn't
already wreaked by randomized advice. \ So for example, I'll conjecture that
just as the class $\mathsf{MA/rpoly}$\ is strictly contained in $\mathsf{ALL}%
$,\footnote{Indeed $\mathsf{MA/rpoly}=\mathsf{MA/poly}$; that is, we can
replace the randomized advice by deterministic advice.} so too its quantum
analogue $\mathsf{QMA/qpoly}$ is strictly contained in $\mathsf{ALL}$. \ I
might be proven wrong, but that's the whole point!

\section*{Conclusion}

For almost a century, quantum mechanics was like a Kabbalistic secret that God
revealed to Bohr, Bohr revealed to the physicists, and the physicists revealed
(clearly) to no one. \ So long as the lasers and transistors worked, the rest
of us shrugged at all the talk of complementarity\ and wave-particle
duality,\ taking for granted that we'd never understand, or need to
understand, what such things actually meant. \ But today---largely because of
quantum computing---the Schr\"{o}dinger's cat is out of the bag, and all of us
are being forced to confront the exponential Beast that lurks inside
our\ current picture of the world. And as you'd expect, not everyone is happy
about that, just as the physicists themselves weren't all happy when
\textit{they} first had to confront it the 1920's.

Yet this unease has to contend with two traditions of technical results: the
first showing that many of the obvious alternatives to quantum mechanics are
nonstarters;\ and the second showing that quantum mechanics isn't
\textit{quite} as strange as one would na\"{\i}vely think. \ Both traditions
are decades old: the first includes Bell's theorem \cite{bell} and the
Kochen-Specker theorem \cite{ks}, while the second includes Holevo's theorem
\cite{holevo}\ and the results of decoherence theory. \ But theoretical
computer scientists come to quantum mechanics with their own set of
assumptions (some would say prejudices), so in this abstract I've tried to
indicate how\ they, too, might eventually be shoved into the vast quantum
ocean, which isn't \textit{that} cold once one gets used to it.

\bibliographystyle{plain}
\bibliography{thesis}

\end{document}